# Observation of two thresholds leading to polariton condensation in 2D hybrid perovskites

*Laura Polimeno*[1,2]*, *Antonio Fieramosca*[1,2]*, *Giovanni Lerario*[1]†,
*Marco Cinquino*[1,2], *Milena De Giorgi* [1], *Dario Ballarini* [1], *Francesco Todisco*[1],
*Lorenzo Dominici* [1], *Vincenzo Ardizzone*[1,2], *Marco Pugliese*[1,2],
*Carmela T. Prontera*[1], *Vincenzo Maiorano*[1], *Giuseppe Gigli* [1,2],
*Luisa De Marco*[1]†, *Daniele Sanvitto*[1]

[1]CNR Nanotec, Institute of Nanotechnology, via Monteroni, 73100, Lecce, Italy.
[2]Dipartimento di Matematica e Fisica "Ennio de Giorgi", Universitá del Salento, Campus Ecotekne, via Monteroni, Lecce, 73100, Italy.

## Abstract

Two dimensional (2D) perovskites are promising materials for photonic applications, given their outstanding nonlinear optical properties, ease of fabrication and versatility. In particular, exploiting their high oscillator strength, the crystalline form of 2D perovskites can be used as excitonic medium in optical microcavities, allowing for the study of their optical properties in the strong light-matter coupling regime. While polariton condensation has been observed in different materials at room temperature, here we observe for the first time two distinct threshold processes in a 2D perovskite, a material that has never shown spontaneous phase transition up to now. In particular, we demonstrate lasing from the bi-exciton state which contributes to populate the lower polariton branch and, at higher excitation powers, eventually leads to the formation of a polariton condensate. The emission linewidth narrowing and a spatial coherence over $50 \times 50\ \mu m^2$ area are the smoking gun, the formation of a quantum coherent state in 2D hybrid perovskite. Our results not only show the formation of a polariton condensate in 2D perovskites but they are also crucial for the understanding of the physical mechanisms that leads to coherent phase transition in perovskite-based polariton microcavities.

---

*These authors contributed equally to this work.
†mail: giovannilerario86@gmail.com, luisa.demarco@nanotec.cnr.it

# 1 Introduction

In the last decade, hybrid organic-inorganic 2D perovskites have attracted considerable interest because of their intriguing optical properties[1,2,3] and better moisture stability than their bulk counterpart[4]. In particular, these materials show a natural quantum well-like structure formed by an inorganic layer of $(PbX_6)^{2-}$ octahedra (in which X is an halogen), sandwiched between bilayers of intercalated alkylammonium cations. In this type of structure, the organic cation behaves as a potential barrier, while the excitons are confined within the inorganic layer[5,6]. The nature of inorganic and organic components defines the quantum well architecture that can be opportunely engineered by changing the chemistry of the synthetic process in order to tailor the optical bandgap and the exciton confinement[7,8].

Recently, Chong *et al.*[9] have highlighted the difficulties in achieving optical gain in a bare single crystal of 2D perovskite, predicting a theoretical bi-exciton amplified spontaneous emission (ASE) threshold of *c* 1.4 $mJ/cm^2$ that goes beyond the crystal damage threshold.

As we demonstrate in this paper, this limit can be overcome embedding a single crystal in an optical microcavity exploiting the strong light-matter coupling regime, i.e. the formation of exciton-polaritons[3]. These bosonic quasi-particles—arising from the strong coupling between excitons and photons—have unique properties inherited from their bare components: the polariton mass is three order of magnitude lighter than the one of the exciton thanks to the photonic component[10], whereas the polariton Kerr nonlinerities—that are up to four orders of magnitude higher than those of standard nonlinear optical media[11]—are inherited from the excitonic component. These peculiar properties make microcavity polariton an ideal candidate for integrated photonic circuits and electro-optical devices[12,13,14]. Furthermore, many fascinating physical phenomena related to quantum fluids, such as polariton condensation[15,16], superfluidity[17,18] and quantized vortices dynamics[19], have been studied both in inorganic and organic polariton devices. In particular, 3D ($ABX_3$)[20] and 2D[21,22,23,24] perovskites are widely used as active medium in optical microcavities, leading to the formation of a strong light-matter system stable at room temperature. Moreover, polariton lasing and condensation have been described by Su *et al.*[25,26] in 3D $CsPbX_3$ perovskites, although exciton-polariton condensation in 2D perovskite has never been observed. As pointed out by Schlaus *et al* [27], it is not trivial to unambiguously distinguish polariton condensation from lasing emission in materials with multiple exciton levels, such as in lead halide perovskites, due to the presence of competitive population/depopulation kinetics.

Here, we investigate the optical properties of the 2D single-crystal perovskite polariton microcavity at cryogenic temperature (4K) demonstrating the presence of two different coherent states forming at two distinct power thresholds: a first lasing action from the bi-exciton state contributes to the subsequent collapse of the polariton population to a macroscopically coherent polariton condensate, characterized by a second threshold at higher pump powers. Our results bring out the complexity of kinetics in perovskite-based systems and pave the way for the study of macroscopic quantum coherent states in 2D perovskites-based polariton microcavities.

## 2 Results and Discussion

The phenethylammonium lead iodide perovskite ($C_6H_5(CH_2)_2NH_3)_2PbI_4$) (PEAI) single crystals are synthesized by anti-solvent vapor assisted crystallization method on a glass substrate[28]. Millimeter-sized flakes having a thickness varying from few to ten micrometres are mechanically exfoliated to improve the surface quality of crystals and to obtain the desired thickness (see Experimental Section for further details). The photoluminescence is recorded using a 50 fs pulsed excitation centered above the band gap (2.640 $eV$) in reflection configuration. In the Section I of the SI (Figure S1), we report the reflectance and emission spectra of a PEAI single crystal at 120 $K$. The emission peak is centered at 2.364 $eV$ with 50 $meV$ full width half maximum (FWHM). In the reflectance spectrum we observe a single peak at 2.366 $eV$, exhibiting a typical minimal Stokes shift of the excitonic emission[29].

At 4 $K$ (Figure 1a), the photoluminescence and reflection spectra are more structured, showing the presence of three different peaks. In the emission spectrum at the lowest incident fluence (18.34 $\mu J/cm^2$), the peaks centered at $E_3$ = 2.341 $eV$ and $E_2$ = 2.330 $eV$ are associated to two electron-phonon replicas of the lowest exciton state[30], while the peak at $E_1$ = 2.300 $eV$ is attributed to Frenkel defects[9]. When increasing the pump power, a narrow peak appears in the emission spectrum at $E_{biex}$ = 2.290 $eV$ (Figure 1b), which is usually attributed to the bi-exciton state[9,31,32], despite its true nature is yet unclear. At low pump fluence, the broad defect peak at $E_1$ = 2.300 $eV$ is energetically overlapped to the bi-exciton emission, that is weaker than Frenkel defects emission. At increasing pump powers, the bi-exciton emission becomes visible and its intensity continuously increases, i.e. accumulates population, while the emission of all the other states saturates (inset in Figure 1b), as previously reported in Ref[9]. Our measurements are obtained on the bare material outside any microcavity resonator, however the trends of the photoluminescence signals with respect to pumping power are fully consistent with the ones reported in Booker *et al.*[33] (see SI, Figure S2), where the authors claim microcavity bi-exciton lasing. We also report a bi-exciton emission linewidth even narrower than the one of the lasing signal reported in Ref[33], despite, in our case, the material is definitely not in the lasing regime. Therefore, in our opinion, the signal observed in Ref.[33] should be interpreted as incoherent emission filtered by the optical microcavity rather than a lasing signal (see Section II of the SI for further details).

We further investigate PEAI perovskite embedded in an optical microcavity. In this case, single crystals are grown on a Distributed Bragg Reflector (DBR, Figure S3) followed by 80 $nm$-thick silver layer evaporated on top of the perovskite crystal (Figure 1c). The microcavity quality factor is $Q > 1000$ (see Section IV, Figure S4). The photoluminescence energy dispersion (Figure 1d) shows the lower polariton branches of a 3 $\mu m$-thick single crystal of PEAI. Multiple lower polariton branches are due to the strong coupling (see SI, Figure S5) of the lowest-energy excitons ($E_{ex}$ = 2.355 $eV$) with different optical modes, according to the microcavity free spectral range[7]. In Figure 1d, the blue dashed lines are the theoretical lower polariton branches resulting from the strong coupling between the exciton transition (red dashed line) and the TE and TM polarized cavity modes (green dashed lines) with a Rabi splitting $\Omega \cong 110$ $meV$ (see Section V for the detailed analysis).

A very useful insight on these light-matter levels and their related phenomenology can be obtained studying their non-linear response at increasing pump fluences. All the optical measurements are performed in reflection configuration at cryogenic temperature, exciting and collecting the signal from the DBR side.

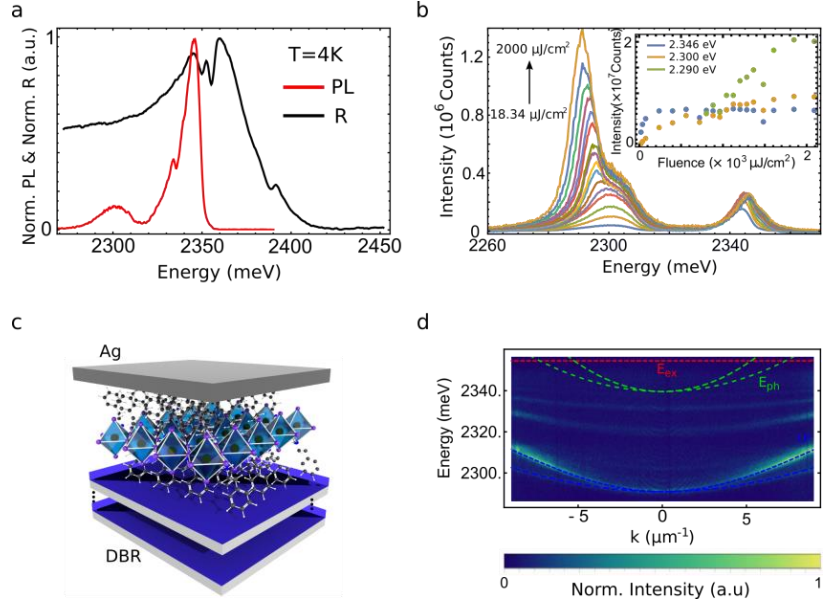

Figure 1: a) Normalized Photoluminescence (red line) and normalized reflectance (black line) spectra of a 70 $nm$-thick single crystal of PEAI at 4 K. Three different peaks are observed ($E_1$ = 2.300 $eV$, $E_2$ = 2.330 $eV$ and $E_3$ = 2.341 $eV$) in the photoluminescence spectrum. b) Power dependent emission of the single crystal PEAI at 4 K. When the pump fluence increases, a narrow peak appears at $E_{biex}$ = 2.290 $eV$ and it is attributed to the bi-exciton emission. Inset: Comparison of the intensities peaks of the three different emission peaks as function of the pump fluence. The intensity of the bi-exciton state (black dots) keeps increasing while the two exciton states saturate (blue dots for peak at $E_1$ and orange dots for peak at $E_2$) at increasing pump power. c) Sketch of the optical microcavity sample. PEAI flakes are embedded in an optical microcavity made by a bottom DBR and a top silver mirror. d) Energy dispersion of the emission of a 3 $\mu m$ single crystal embedded in a planar microcavity, in which multiple lower polariton branches are visible. The blue dashed line is the theoretical dispersion of the eigenstates of the coupled system resulting from the strong coupling between the excitonic transition at $E_{ex}$ = 2.355 $eV$ (red dashed line) and the optical modes in green dashed lines (the two dispersions correspond to the TE and TM polarized modes).

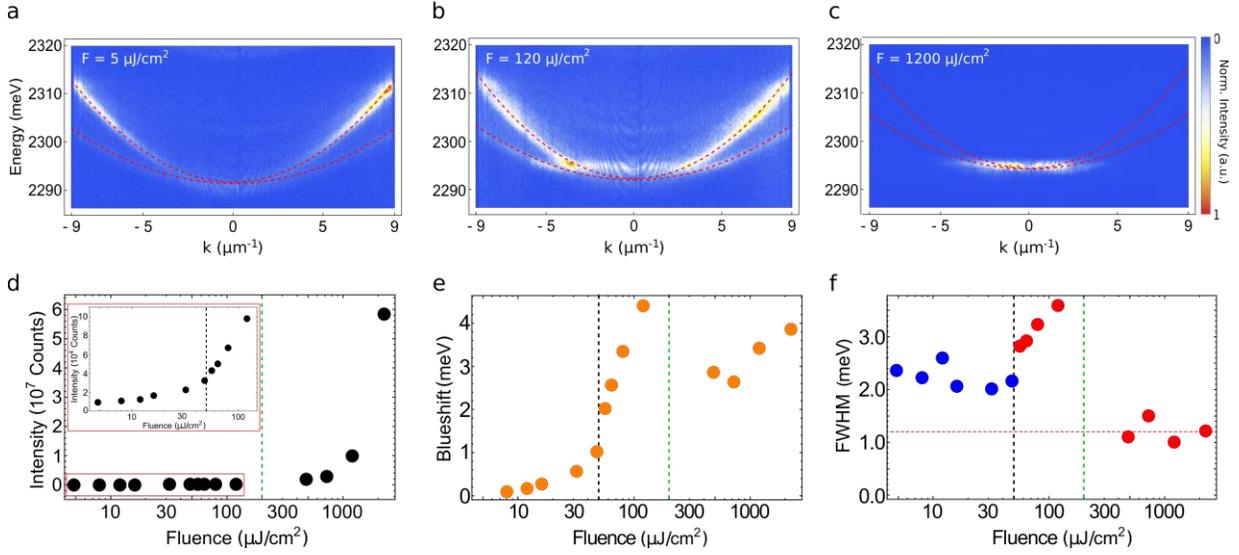

Figure 2: a), b), c) Energy Vs momentum emission intensity maps for three different incident pump fluences, exciting a 3 $\mu m$-thick perovskite single-crystal embedded in a planar microcavity (50-fs pulsed excitation at 2.64 $eV$). The red dashed lines are a guide to the eye for the TE and TM polariton modes. The black dashed line indicates the bottom energy of the lower polariton branch at very low excitation power. a) At 5 $\mu J/cm^2$, the emission is intense at high-energy and high in-plane wavevectors. b) Increasing the incident fluence up to 50 $\mu J/cm^2$, the bi-exciton state starts lasing and we observe an intense peak appearing at about $E = 2.295\ eV$. c) Further increasing the pump fluence, the whole emission collapses to the bottom of the lower polariton branch ($k \cong 0\ \mu m^{-1}$), generating a polariton condensate. d) Integrated intensity of the emission as function of incident pump fluences. Inset: Zoom for weak excitation intensities. Two different thresholds are clearly visible, the first one at $F = 50\ \mu J/cm^2$ (black dashed line) indicates the bi-exciton lasing activation, and the second one at $F = 200\ \mu J/cm^2$ (green dashed line) is associated to the formation of the polariton condensate at the lowest energy state. e) Energy blueshift of the emission as a function of incident fluences. At the onset of the polariton condensate, the population collapses to the bottom of the polariton dispersion. f) The FWHM of the bi-exciton lasing emission increases at increasing pump power and it is broader than the FWHM of the polariton condensate at the lowest energy state. e) Energy blueshift of the emission as a function of incident fluences. At the onset of the polariton condensate, the population collapses to the bottom of the polariton dispersion. f) The FWHM of the bi-exciton lasing emission increases at increasing pump power and it is broader than the FWHM of the polariton dispersion below the lasing threshold (blue dots). For pump fluences above $F = 200\ \mu J/cm^2$ (green dashed line), the FWHM narrows due to the onset of the polariton condensate. The red dashed line indicates the optical setup resolution (1.2 $meV$).

Figure 2a, 2b and 2c show the power dependence of the polariton emission dispersion of the perovskite microcavity under a 50 $fs$ pulsed laser (Gaussian spot size with FWHM= 5 $\mu m$). At low fluence $F$ = 5 $\mu J/cm^2$ (Figure 2a), the emission is intense at high in-plane wavevectors of the lower polariton branch (LPB). Increasing the incident pump power, we observe a first threshold at about $F$ = 50 $\mu J/cm^2$ where an intense emission peak appears at about $E$ = 2.295 $eV$, a few meV above the bottom of the LPB. We attribute this peak to the onset of a lasing signal, given the narrowing of the linewidth and the nonlinear increase of the emission intensity (Figure 2d and 2f). Laser linewidth broadening and peak blueshift are observed at increasing pump power. This behaviour leads to the conclusion that this signal is related to the bi-exciton lasing action rather than to the polariton condensate emission since the system is strongly out of equilibrium, as confirmed by the lasing energy—which is higher than the bottom of the LPB. The blueshift of the lasing emission could be due to phase space filling effects or to interaction between bi-excitons. Moreover, as there is no thermalization (see SI Figure S5), the lasing action is given by the cavity-induced gain at the uncoupled bi-exciton state, which emission energy is resonant to a specific microcavity mode ($E_{biex}$ = 2.290 $eV$). Considering the population accumulation to the bi-exciton state observed in the out-of-cavity perovskite layer (Figure 1b) at increasing pump powers, the fact that the bi-exciton state is the first one to achieve the lasing threshold (i.e. the population inversion) is also theoretically expected[9]. The blueshift of the lasing state (Figure 2e, points in between the black and the green dashed lines) does not relate to a blueshift of the lower polariton dispersion, and the bi-exciton emission peak is broader than the polariton linewidth (about 2.3 $meV$, Figure 2f). This demonstrate that both the bi-exciton emission linewidth and the energy

blueshift are independent from the polariton states. To further confirm such interpretation we observe at higher fluences ($F = 200~\mu J/cm^2$), a second phase transition to a higher coherent state at the energy minimum of the LPB associated with a dramatic increase of the emission intensity (Figure 2d). As it can clearly be observed in Figure 2c, the whole population collapses to the lower energy state, with almost no signal from higher energy states—also the bi-exciton lasing action turns off—as predicted in case of bosonic stimulation. The population energy distribution in this new regime is very different from the one of the bi-exciton lasing (see SI, Figure S6b). When the bosonic condensate is established at the bottom of the lower polariton branch, the FWHM narrows indicating the increase of the coherence time, and its spectral width goes below our setup resolution ($1.2~meV$, red dashed in Figure 2f). In this regime, the energy shift of the emission (Figure 2e) is consistent with a partial saturation of the coupled exciton state[34].

These results also demonstrate that the physics of polariton condensation in such complex materials with many exciton levels, cannot be solely explained by considering the physics of a two level system.

We point out that no photo-degradation of the sample have been observed with pump fluence up to $300~\mu J/cm^2$ and for an exposure of about one hour. The polariton condensate forms exclusively when the energy of the bottom of the polariton mode is close to the energy of the bi-exciton lasing emission, otherwise we only observe the perovskite photodegradation when keep increasing the pump power.

To further corroborate the formation of a polariton condensate, we perform first-order spatial coherent measurements $g^1(r, -r)$[17]—making use of a Michelson interferometer along the detection line—the Gaussian spot size is 50x50 $\mu m^2$ (see Experimental Section for further informations). A back-reflector flips the real space emission map on one arm of the interferometer in a centro-symmetrical way so that the resulting image detected on the CCD camera is the superposition of the reference signal (r) and the inverted one (-r). Figure 3 shows the $g^1(r, -r)$ signals in the linear regime (Figure 3a at $F = 5\ \mu J/cm^2$) and above the second threshold (Figure 3b at $F = 600\ \mu J/cm^2$). In the latter case, a typical pattern of interference is clearly visible for the whole size of the condensate, demonstrating long range coherence over 50x50 $\mu m^2$. Note that for the bi-exciton lasing action, the real space emission is characterized by a strong inhomogeneity (see SI, Figure S7), preventing the possibility to observe a clear $g^1$ signal over a large area.

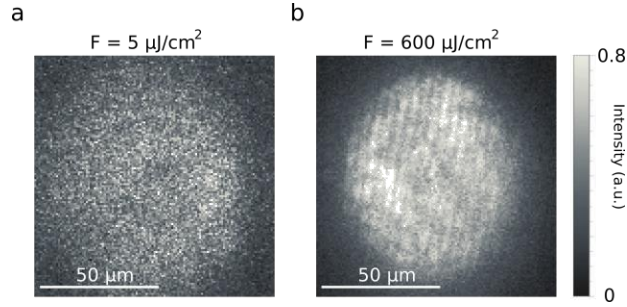

Figure 3: a),b) Real space emission maps obtained superimposing the two images split along the two arms of a Michelson interferometer. a) In the linear regime, no interference fringes appears in the coherence map. b) Above the condensation threshold, a macroscopic coherent state arises and interference fringes appear over 50x50 $\mu m^2$, which is also the excitation spot size.

Only high quality 2D crystals show two distinct thresholds, while we observe different behaviours studying several kind of samples with different degree of disorder. Specifically, polariton condensation does not build up when the perovskite crystals include structural defects. These defects are lasing at very low pump power (lasing threshold observed also at 0.4 $\mu J/cm^2$ pump fluence), they are localized in space and have disperse emission energy (see SI, Figure S8). In this case, there is no observation of bi-exciton lasing or polariton condensation up to the maximum pump power before material damaging. In addition, when structural defects are present, we have also noted a reduction of the perovskite layer damaging threshold compared to the one observed in high quality 2D crystals (photodegradation can also occur for pump fluences close to 1 $\mu J/cm^2$).

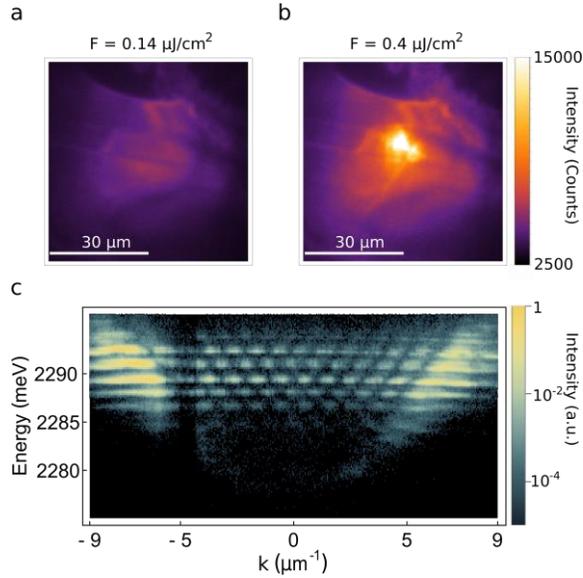

Figure 4: a), b) Real space emission intensity maps in presence of a 0$D$ defect for two different pump fluences. The non-linear increase of the confined-states emission intensity demonstrate the lasing action (see SI Section VII for further details). Gaussian excitation spot size of 50x50 $\mu m^2$. c) Energy Vs momentum emission of the confined states for $F = 0.4$ $\mu J/cm^2$.

On the other hand, optical confinement phenomena take place at the position of some particular structural defects of the perovskite crystals[35,36], similarly to what have been reported in 3D $CsPbBr_3$ perovskites[37]. When a defect induces a confining potential in the device plane, lasing action from multiple high-energy confined states is observed in the real space maps (Figure 4a,b) at the defect position and, as well, in the energy dispersion map (Figure 4c). Note that the lasing action from these confined states is always resonant to the defect energy band, even when varying the cavity detuning, therefore the stimulated emission can also be activated at the lowest-energy confined state (see SI Figure S9, Figure S10).

# 3 Conclusion

We have demonstrated for the first time both lasing from the bi-exciton state and the subsequent generation of a polariton condensate, in hybrid organic-inorganic 2D layered perovskite crystals. Furthermore, to our knowledge, this is the most clear evidence of polariton condensation, in an organic or hybrid material, which is unambiguously different from lasing effect and can here be clearly distinguished. The polariton condensate forms at higher excitation power with respect to a bi-exciton lasing action. The collapse of the polariton population toward the minimum of the lower polariton dispersion turns off the bi-exciton emission and the total emission intensity abruptly increases building up a long-range spatial coherent state.

These experimental results demonstrate that such materials cannot be described using a simple two-level-system model since many energetically competitive phenomena are at stake.

This work sheds new lights on the physics of hybrid and organic crystals in optical confined systems, giving the possibility to investigate quantum coherent states based on organic-inorganic 2D layered perovskites.

# Experimental Section

# 4 Synthesis of 2D perovskite flakes

1M PEAI solution is prepared in a nitrogen-filled glovebox by dissolving $PbI_2$ and phenethylammonium iodide (1:2 molar ratio) in $\gamma$-butyrolactone and stirring at 70°$C$ for 1 hour. 2D perovskite single crystals are synthesized using an anti-solvent vapor-assisted crystallization method. 5 $\mu$l of perovskite solution are deposited on top a sputtered DBR (or a glass substrate) and covered with a glass coverslip. These substrates and a small vial containing 2 ml of dichlorometane (anti-solvent) are placed inside a bigger Teflon vial which is closed with a screw cap and left undisturbed for 12 hours. After this time, millimeter-sized crystals appear in between the two substrates and their thickness vary from few to tens of micrometers. Using SPV 224PR-M Nitto Tape or PDMS, mechanical exfoliation is carried out on the perovskite flakes in order to obtain single crystals having the desired thickness.

# 5 Microcavity Sample Fabrication

The DBR is made by seven pairs of $TiO_2/SiO_2$ ($53\ nm$/$89\ nm$) deposited by radio-frequency (RF) sputtering process—in an Argon atmosphere under a total pressure of $6 \cdot 10^3$ mbar and at RF power of $250\ W$ —on top of a $170\ \mu m$ glass substrate. The perovskite single crystals are grown on top of the DBR (see above) and a $80\ nm$-thick layer of silver is thermally evaporated on top of the structure (deposition parameters: current = $280\ A$, deposition-rate = $3$ Å/s).

# 6 Optical measurements

Figure 5 shows the sketch of the setup used for the optical measurements. The measurements are performed at cryogenic temperature (T=4K) using a 50-fs pulsed laser at $470\ nm$ ($10\ kHz$ repetition rate) for emission measurements and a white light Xenon lamp for the reflectance spectra.

Through a first detection path, the image on the back focal plane of the detection objective is projected on a spectrometer entrance slit. The spectrometer is coupled to an enhanced CCD camera for the detection of the polariton energy dispersion. A second detection path is used for the spatial intensity maps and spatial coherence measurements ($g^1(r, -r)$). In this latter case, the single-crystal emission signal enters into a Michelson interferometer; a movable retroreflector is placed along one of the two arms of the interfermeter in order to inverts the image in a centro-symmetrical configuration. The image containing the superposition of the two interferometer arms is then sent to a CCD camera. The overall magnification is $30X$.

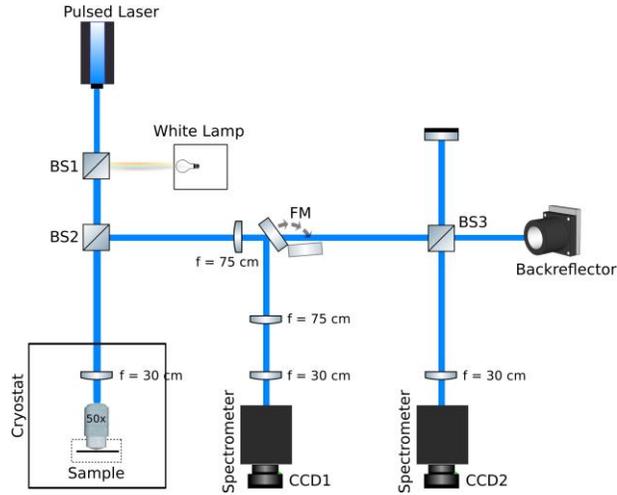

Figure 5: Sketch of the optical setup.

## Acknowledgments

The authors acknowledge Paolo Cazzato for technical support, Iolena Tarantini for the metal evaporation and Carlo Giansante for useful discussions. We acknowledge the project "TECNOMED - Tecnopolo di Nanotecnologia e Fotonica per la Medicina di Precisione", (Ministry of University and Scientific Research (MIUR) Decreto Direttoriale n. $3449$ del $4/12/2017$, CUP $B83B17000010001$).

## I. PHOTOLUMINESCENCE AND REFLECTION SPECTRA AT 120 K

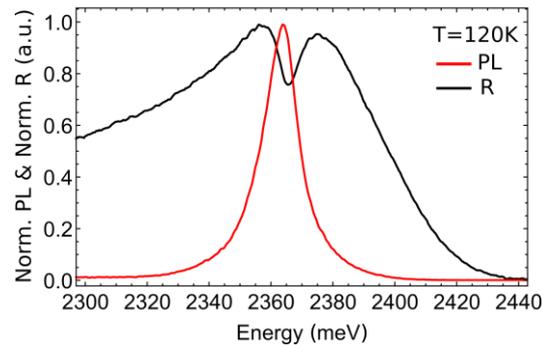

Figure S1: Normalized Photoluminescence (red line) and normalized reflectance (black line) spectra of a 70 $nm$-thick single crystal of PEAI at 120 K. The emission peak is centered at 2.360 $eV$ with a FWHM of 50 $meV$ and a single peak at 2.360 $eV$ is visible in the reflectance spectrum.

## II. EMISSION INTENSITY AS A FUNCTION OF PUMP FLUENCE

Figure S2 shows the the emission of the bi-exciton state (blue dots) and the exciton state at $E = 2.310\ eV$ (orange dots) of the bare perovskite layer, as a function of the pump fluence. Our results are fully comparable with the behaviour reported in Ref.[29], where bi-exciton lasing is claimed. In particular, in Booker *et al.*[33] manuscript, not only a clear narrowing of the emission linewidth with respect to the bare bi-exciton spectrum is not observed — that also exclude even an ASE signal — but also the authors does not provide any coherence measurement to support their claim. For these reasons, we suggest to interpret their observation as standard biexction photoluminescence filtered by an optical microcavity.

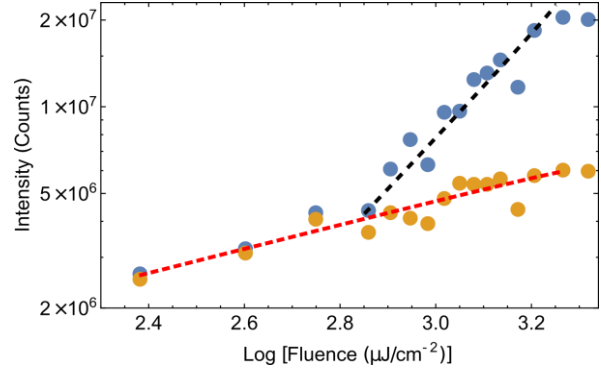

Figure S 2: Comparison between integrated intensity of the bi-exciton state (Blue dots) at $E_{biex}$ = 2.290 $eV$ and the exciton state (Orange dots) at $E$ = 2.310 $eV$, plotted in logarithmic scale as reported in Booker *et al.*[33].

### III. TRANSMISSION SPECTRUM OF THE DBR

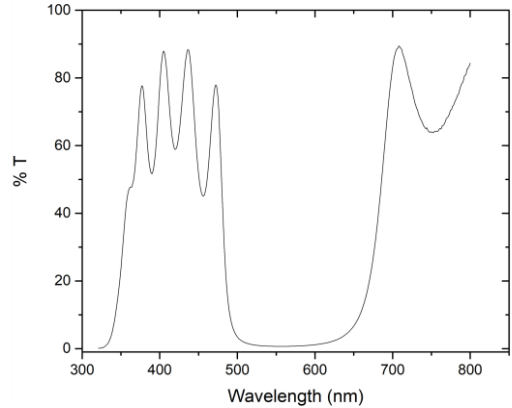

Figure S 3: Trasmission spectrum of the bottom DBR.

### IV. Q-FACTOR

Considering a Lorentian fitting of the polariton mode, we can extract a FWHM of about 2.3 $meV$. The *Q-factor* of a cavity with Lorentzian spectrum is $\frac{E}{\Delta E}$. In the present case we obtain $\frac{E_{pol}}{\Delta E_{pol}} \cong 1000$ for the polariton mode. Since the *Q-factor* is defined on the empty cavity mode, the $\frac{E_{pol}}{\Delta E_{pol}}$ is a lower limit of the *Q-factor* considering the perovskite dissipation.

The polariton mode under investigation is far from the exciton transition, therefore the exciton dissipation have weak influence on the mode broadening.

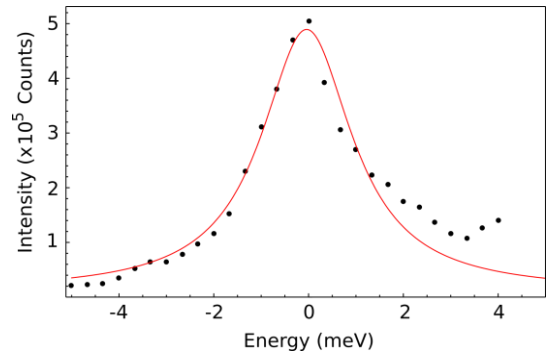

Figure S 4: Experimental data (dots) of the polariton emission line and Lorentzian fit (red line) with $FWHM = 2.33\ meV$.

## V.  STRONG-COUPLING

According to the two coupled oscillator system, we evaluate the polariton energy dispersion as a function of the photon energy, considering the experimental data of the all lower polariton branches (Figure S3a) and the exciton energy:

$$(\omega_{exc} - \omega)(\omega_{ph} - \omega) = \left(\frac{\Omega}{2}\right)^2 \quad (1)$$

where $\Omega$ is the Rabi energy splitting, $\omega_{exc}$ = 2.355 $eV$ is the exciton energy — extracted from reflection spectrum — and $\omega_{ph}$ is the photon energy.

In Figure S4a, experimental multiple lower polariton branches are plotted as function of the photon energy with different colors for the whole range of the in-plane momentum k. In Figure S4b, the experimental value of the lower polariton energies are plotted in blue, the black line is the lower polariton energy resulting from the fitting with $\Omega \cong 110$ $meV$. The orange line is the upper polariton energy resulting from the two coupled oscillator system — the upper polariton branches are not experimentally visible due to dissipation related to the perovskite high-energy states.

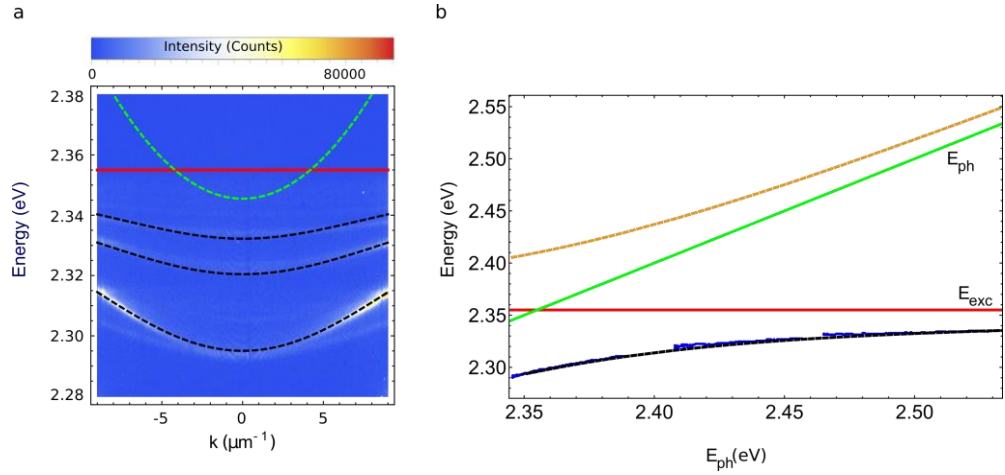

Figure S5: a) lower polariton dispersions (LPBs) maps. The red line is the exciton energy, the green line is the photon dispersion for the lowest polariton mode. The black dashed lines are the values of the fitting of the LPBs considering a two coupled oscillator model with Rabi splitting $\Omega \cong 110\ meV$
b) Calculated Upper polariton (orange line) and lower polariton (black line) eigenergies as a function of the photon energy. The blue points are the experimental data of the dispersion maxima for the three LPBs. The green line is the bare photon dispersion and the red line is exciton transition at $E_{exc} = 2.355\ eV$.

## VI. FLUENCE DEPENDENCE OF POPULATION DISTRIBUTION

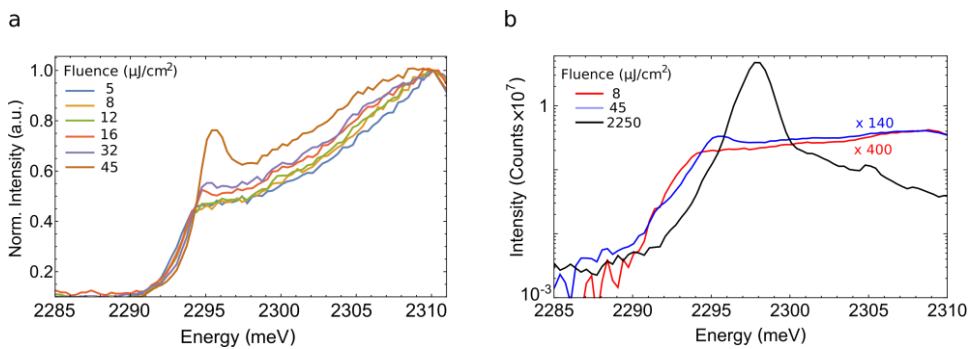

Figure S6: a) Comparison between integrated emission spectra (normalized to the emission signal at 2310 meV) at different pump fluences, up to the bi-exciton lasing threshold. There is no evidence of thermalization effect. b) Comparison between integrated emission spectra at different pump fluences. The polariton condensate emission distribution (black line) is substantially different from the bi-exciton lasing (blue line) and spontaneous photoluminescence signals (red line).

## VII. REAL SPACE EMISSION OF BI-EXCITON LASING

The real space map of the bi-exciton lasing (Figure S5) shows an inhomogeneous and fragmented emission. Because of the fragmentation, there is no spatial overlap between the mirrored images necessary for the *g(1)* coherence measurement.

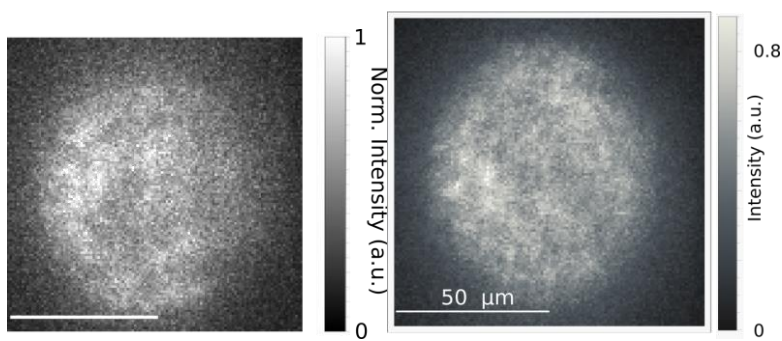

Figure S 7: Real space emission map at bi-exciton lasing threshold. The bi-exciton lasing is inhomogeneus in space. The scale bar is $50 \ \mu m$

## VIII. LASING FROM DEFECTS

In sample areas where defect are present we observe lasing from these defects instead of bi-excioton lasing and polariton condensation (Figure S7). Depending on the defect, the stimulated emission is reached at different excitation power, ranging from very low pump fluence ($F = 0.4\ \mu J/cm^2$), as shown in Figure S9a, to pump fluence of $F = 4\ \mu J/cm^2$, as shown in Figure S9b.

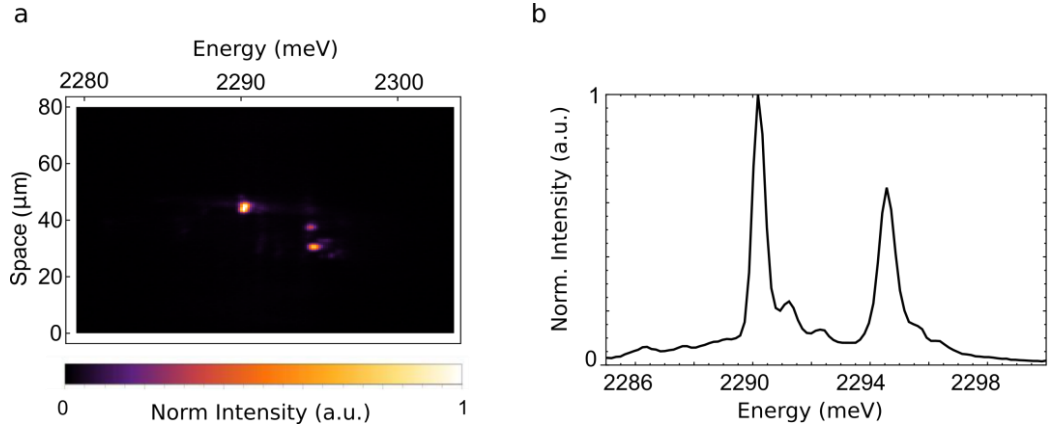

Figure S 8: a) Lasing emission spectrum from localized defects along the sample. b) Same emission spectrum integrated in space. The pump is a 50-fs pulsed laser at $2.64\ eV$.

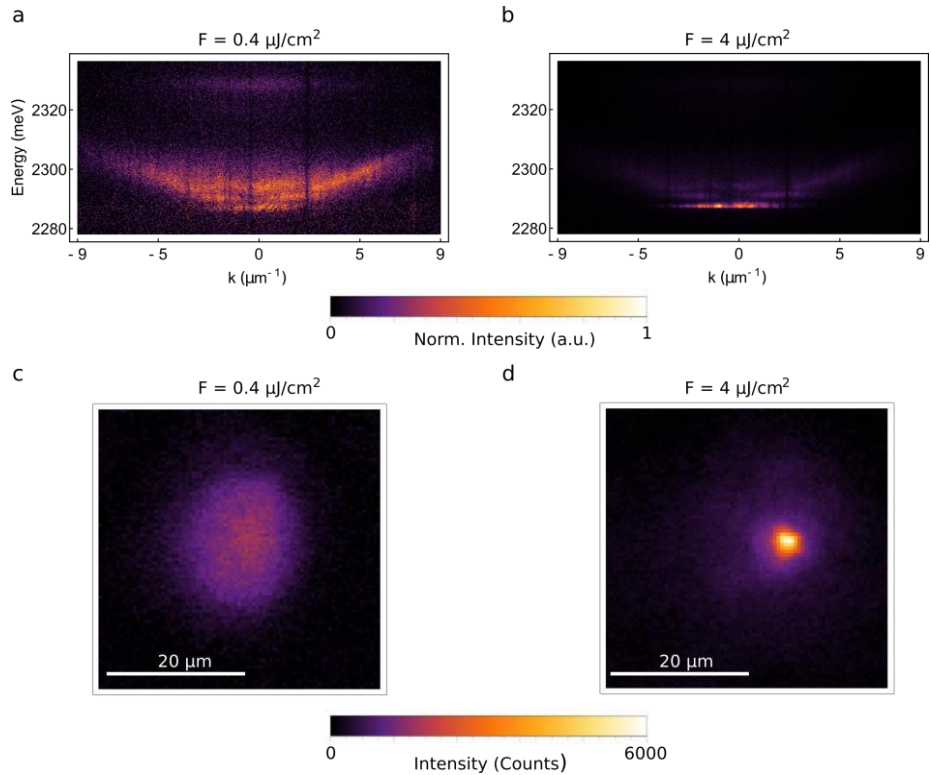

Figure S 9: a), b) Emission energy dispersion for a) $F = 0.4\ \mu J/cm^2$ and b) $F = 4\ \mu J/cm^2$ incident fluences in an area with an optical defect. The coherent emission coming from the lower energy-state is visible at high pump fluence. c),d) Real space emission for low ($F = 0.4\ \mu J/cm^2$) and high ($F = 4\ \mu J/cm^2$) pump fluence. The stimulated emission is localized in real space at the potential well position.

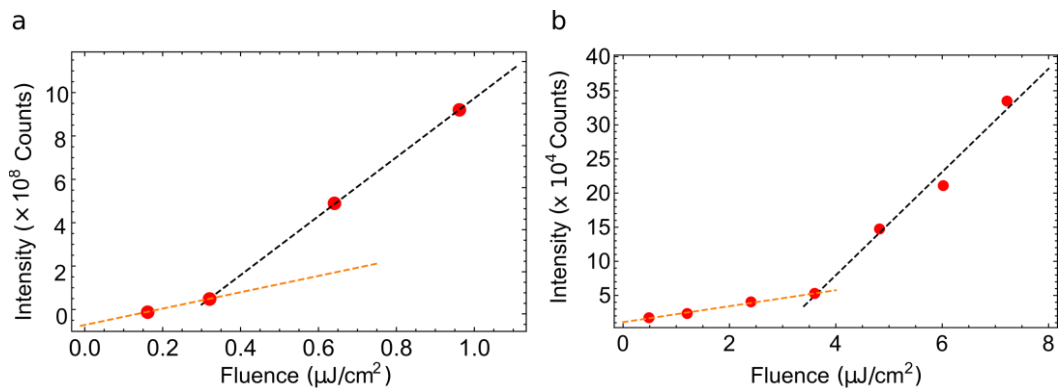

Figure S 10: Emission intensity of a) the lasing defect reported in Figure 4 of the main text and of b) the lasing defect reported in Figure S8, as a function of the pump fluence.